\documentclass[prd,twocolumn,superscriptaddress,showpacs,nofootinbib,preprintnumbers]{revtex4}

\usepackage{amsmath}
\usepackage{amsfonts}
\usepackage{graphicx}
\usepackage{dcolumn}

\def\be{\begin{equation}}
\def\ee{\end{equation}}
\def\ba{\begin{eqnarray}}
\def\ea{\end{eqnarray}}
\def\bs{\begin{subequations}}
\def\es{\end{subequations}}

\newcommand{\rd}{{\rm d}}

\newcommand{\vp}{\varphi}

\usepackage{color}

\begin{document}

\title{Reconstruction of general scalar-field 
dark energy models}

\author{Shinji Tsujikawa}
\affiliation{Department of Physics, Gunma National College of
Technology, Gunma 371-8530, Japan}
\date{\today}

\begin{abstract}

The reconstruction of scalar-field dark energy models is 
studied for a general Lagrangian density $p(\phi, X)$,
where $X$ is a kinematic term of a scalar field $\phi$.
We implement the coupling $Q$ between dark energy and 
dark matter and express reconstruction equations
using two observables: the Hubble parameter $H$
and the matter density perturbation $\delta_m$.
This allows us to determine the structure of corresponding 
theoretical Lagrangian together with the coupling $Q$
from observations. We apply our formula to several 
forms of Lagrangian and present concrete examples
of reconstruction by using the recent Gold dataset of supernovae
measurements. This analysis includes a generalized ghost condensate 
model as a way to cross a cosmological-constant boundary
even for a single-field case.

\end{abstract}

\maketitle

\section{Introduction}

Observations suggest that our universe has entered a stage of an
accelerated expansion with a redshift $z \lesssim 1$ \cite{Perl}.
This has been supported by a number of recent astrophysical data
including supernovae (SN) Ia \cite{SN}, Cosmic Microwave
Background (CMB) \cite{CMB} and large-scale structure \cite{SDSS}.
These observations show that about 70\% of the total energy
density of the present universe consists of dark energy
responsible for an accelerated expansion (see Refs.\,\cite{obser}
for recent works). The simplest candidate
of dark energy is a cosmological constant, but this scenario
suffers from a severe fine-tuning problem of its energy scale from
the viewpoint of particle physics \cite{review}. Hence it is natural to
pursue alternative possibilities to explain the origin of dark
energy.

So far a wide variety of scalar-field dark energy models has been
proposed-- including quintessence \cite{quin}, k-essence
\cite{Kes}, tachyons \cite{tachyon}, phantoms \cite{phantom} and
ghost condensates \cite{Arkani,PT,KN}. These scenarios are
distinguished from a cosmological constant because of their dynamical
nature of the equation of state of scalar fields. A typical
approach is to predict the evolution of the Hubble parameter 
theoretically and to check the consistency of models 
by comparing it with observations.
An alternative approach is to start from observational data and to
reconstruct corresponding theoretical Lagrangian. The latter is
more efficient to find out best-fit models of dark energy from
observations.

This reconstruction is simple for a minimally coupled scalar field
with a potential $V(\phi)$ \cite{Sta,HT,NC,GOZ}. In fact one can
reconstruct the potential and the equation of state of the field
by parametrising the Hubble parameter $H$ in terms of the
redshift $z$ from the luminosity 
distance $D_L(z)$ \cite{SRSS}. 
This method can be generalized to
scalar-tensor theories \cite{BEPS,EP,Pe}, $f(R)$ gravity
\cite{Capo} and also a dark-energy fluid with viscosity 
terms \cite{CCENO}. 
In scalar-tensor theories a scalar field called dilaton is
coupled to gravity, which  means that  an additional function
$F(\phi)$ exists in front of the Ricci scalar $R$.
If the evolution
of matter perturbations $\delta_m$ is known observationally in
addition to the Hubble parameter $H(z)$, one can even determine
the function $F(\phi)$ together with the potential $V(\phi)$ of
the scalar field \cite{BEPS}.

In this paper we will provide a reconstruction program for a very
general scalar-field Lagrangian density $p(\phi, X)$, 
where $X \equiv -(\nabla \phi)^2/2$ is a kinematic term. 
We shall consider an Einstein-Hilbert action 
with a scalar field $\phi$ coupled to a
non-relativistic barotropic fluid (dark matter) with a coupling 
$Q(\phi)$. This coupled quintessence scenario was proposed in
Ref.~\cite{Luca} as an extension of scalar-tensor theories
with a nonminimally coupled scalar 
field \cite{Lucanon,Uzan,Tchiba,BP,Mata}. 
In fact the scalar-tensor action in
Jordan frame can be transformed to the action in Einstein frame 
with an explicit coupling between the scalar field and the
non-relativistic fluid.

The presence of the coupling $Q(\phi)$ means that the
parametrisation of the Hubble rate $H(z)$ is not sufficient to
determine the structure of theory. We shall use the equation of
sub-Hubble matter perturbations $\delta_m$ recently derived in
Ref.~\cite{Amenphan} for the Lagrangian density 
$p(\phi, X)$ (see also Refs.~\cite{Amenper,ATS}). 
The coupling $Q(\phi)$ is 
determined once we know $H(z)$ and $\delta_m(z)$
observationally. We note that this 
places constraints on the strength of nonminimal couplings
when the coupling $Q$ originates from scalar-tensor theories.

The observations of Sloan Digital Sky Survey 
(SDSS) and Two degree Field (2dF) galaxy 
clustering \cite{gala1,gala2,gala3}
provide the information of matter perturbations.
Galaxy clustering is proportional to matter clustering
on large scales with a constant of proportionality called 
bias. The analysis of galaxy clustering itself does not 
determine the value of bias.
Hence the evolution of matter perturbations
is difficult to be known unless bias is determined 
observationally.
However recent observations using the luminosity 
dependence of galaxy clustering together with 
the halo mass distribution began to provide good data
for the determination of bias \cite{gala2,Sel1,Sel2}.
It is expected that we will be able to determine
the evolution of matter perturbations, $\delta_m(z)$,
from upcoming high-precision observations.
 
The recent SN analysis using the Gold dataset \cite{SN} implies that 
the parametrisation of $H(z)$ which crosses the cosmological-constant
boundary ($w=-1$) shows a good fit to data. This crossing to the
phantom region ($w<-1$) is not possible for an ordinary minimally
coupled scalar field [$p=X-V(\phi)$].
This transition can occur
for the Lagrangian density $p(\phi, X)$ in which 
$\partial p/\partial X$ changes sign from positive to negative,
but we require nonlinear terms in $X$ to realise the 
$w=-1$ crossing \cite{PT,Vikman} \footnote{I thank Alexander Vikman 
for pointing out that the $w=-1$ crossing is hard to be realised only in 
the presence of the terms linear in $X$ \cite{Vikman}.}.
It was shown in Ref.\,\cite{NOT} that such a crossing is possible 
if a (phantom) dark energy fluid is coupled to a non-relativistic 
fluid with a specific coupling. 
This can be also realised in a multi-field system 
with a phantom and an ordinary scalar \cite{multi}.
In this paper we shall show a simple one-field model 
($p=-X+h(\phi)X^2$) crossing 
the cosmological-constant boundary and perform 
the reconstruction of such a theory.

\section{Reconstruction program}

We start with a general action given by
\be
\label{action}
{\cal S}=\int \rd^4 x \sqrt{-g} \left[\frac12
R+p(\phi, X)\right]+
{\cal S}_m(\phi, g_{\mu \nu}) \,, \ee
where $R$ is a scalar curvature,
$p(\phi, X)$ is a general function of $\phi$ and
$X=-(1/2)(\nabla \phi)^2$.
The gravitational constant, $\kappa^2 \equiv 8\pi G$,
is set to be unity.
We assume that the field $\phi$ is coupled
to a barotropic perfect fluid with a coupling $Q(\phi)
\equiv -1/(\rho _{m}\sqrt{-g})\delta {\mathcal{S}}_{m}/\delta
\phi$, where $\rho_m$ is the energy density of the fluid.
We shall use a sign notation $(-, +, +, +)$.

In a flat Friedmann-Robertson-Walker (FRW) spacetime
with a scale factor $a$, 
the field equations for the action (\ref{action}) are
\ba \label{basiceq1}
& &3H^2=\rho_m+2Xp_X-p\,, \\
\label{basiceq2}
& & 2\dot{H}=-\rho_m-p_m-2Xp_X\,, \\
\label{basiceq3} & &
\dot{\rho}_m+3H(\rho_m+p_m)=Q(\phi)\rho_m
\dot{\phi}\,, \ea
where $H=\dot{a}/a$, $X=\dot{\phi}^2/2$, 
$p_X=\partial p/\partial X$ 
and $p_m$ is the pressure density
of the fluid. A dot denotes a derivative with respect to
cosmic time $t$.
In what follows we shall consider
the case of a non-relativistic barotropic fluid,
i.e., $p_m=0$. Eq.~(\ref{basiceq3}) is written
in an integrated form
\ba \rho_m=\rho_m^{(0)} \left(\frac{a_0}{a}\right)^3
I(\phi)\,,
\ea
where 
\ba \label{Idef}
 I(\phi) \equiv \exp \left(\int_{\phi_0}^\phi Q(\phi) \rd
\phi\right)\,. \ea
Here $\rho_m^{(0)}$, $a_0$ and $\phi_0$ are the present values of the
energy density $\rho_m$, the scale factor $a$
and the scalar field $\phi$, respectively. By using the present
ratio $\Omega_{0m}$ of the matter fluid and the Hubble parameter
$H_0$, $\rho_m^{(0)}$ is given by
$\rho_m^{(0)}=3H_0^2\Omega_{0m}$. 
We also define the redshift parameter $z$, 
as $1+z \equiv a_0/a$. 
Then the energy density $\rho_m$ is written as
\ba \label{rhom}
\rho_m=3\Omega_{0m}H_0^2 (1+z)^3 I(\phi)\,. 
\ea

One has $I(\phi)=1$ in the absence of the coupling $Q(\phi)$.
In this case one can reconstruct the structure of theory by 
using Eqs.~(\ref{basiceq1}), (\ref{basiceq2}) and (\ref{rhom})
if the evolution of the Hubble parameter is known 
from observations.
This was actually carried out for an ordinary scalar field 
with a Lagrangian density: $p=X-V(\phi)$ \cite{HT,Sta,NC}.
When the field $\phi$ is coupled to dark matter, we require
additional information to determine the strength of the coupling.
We shall use the equation of matter density perturbations
for this purpose as in the case of scalar-tensor
theories \cite{BEPS,EP}.

The evolution equation for the matter density
contrast, $\delta _{m}\equiv \delta \rho _{m}/\rho _{m}$, was 
recently derived in Ref.~\cite{Amenphan}.
On sub-Hubble scales this is given by
\begin{equation}
\ddot{\delta}_{m}+\left[2H+Q(\phi)\dot{\phi}\right]
\dot{\delta}_m-\frac12
\left[1+\frac{2Q^2(\phi)}{p_X}\right]
\rho_m\delta_m=0\,.
\label{eqmatter}
\end{equation}
This is a very general equation which holds for any scalar-field 
Lagrangian density $p(\phi, X)$ and also for the case in which 
the coupling $Q$ depends on the field $\phi$.
Eq.~(\ref{eqmatter}) can be solved analytically
for scaling solutions \cite{ATS}.
In fact it was shown that 
matter perturbations are suppressed for a phantom ($p_X<0$)
whereas they are not for an ordinary field ($p_X>0$).
Generally we need to specify the Lagrangian density 
$p(X, \phi)$ in order to solve Eq.~(\ref{eqmatter}).
We note that the gravitational potential $\Phi$ is related with 
the matter perturbation $\delta_m$ through the relation 
$\Phi \simeq -(3a^2H^2/2k^2)\Omega_m \delta_m$
on sub-Hubble scales \cite{ATS} ($k$ is a comoving wavenumber).
 
One can rewrite the equations (\ref{basiceq1}), (\ref{basiceq2}) and
(\ref{eqmatter}) by using a dimensionless quantity 
\ba 
r \equiv H^2/H_0^2\,.
\ea
Making use of the relation (\ref{rhom}), we find that
\ba 
\label{re1}
& & p=\left[(1+z)r'-3r\right]H_0^2\,, \\
\label{re2} & &
\phi'^2p_X=\frac{r'-3\Omega_{0m}(1+z)^2I}
{r(1+z)}\,,\\
\label{re3} & &
\delta_m''+\left(\frac{r'}{2r}-\frac{1}{1+z}+
\frac{I'}{I} \right) \delta_m'
\nonumber \\ & &-\frac32 \Omega_{0m}
\left(1+\frac{2I'^2}{\phi'^2p_XI^2} \right)
\frac{(1+z)I\delta_m}{r}=0\,,
 \ea
where a prime represents a derivative with respect to $z$.
Eliminating the $\phi'^2p_X$ term from Eqs.~(\ref{re2})
and (\ref{re3}), we obtain
\ba \label{Iprime}
I'=\frac{I}{4r(1+z)A} 
\left[\delta_m' \pm \sqrt{\delta_m'^2-8r(1+z)AB}\right]\,, 
\ea
where
\ba & & A \equiv \frac{3\Omega_{0m}(1+z)\delta_mI}
{2r[r'-3\Omega_{0m}(1+z)^2I]}\,, \\
& &B \equiv [r'-3\Omega_{0m}(1+z)^2I]A \nonumber \\
&&~~~~~~-\delta_m''
-\left(\frac{r'}{2r}-\frac{1}{1+z}
\right)\delta_m'\,.
\ea
{}From Eq.~(\ref{Iprime}) we require the following condition
\ba \delta_m'^2>8r(1+z)AB\,.
\ea

Once we know $r$ and $\delta_m$ in terms of $z$
observationally, Eq.~(\ref{Iprime}) is integrated to give the
functional form of $I(z)$.  
It is worth mentioning that the function $I(z)$ 
is determined without specifying the Lagrangian 
density $p(\phi, X)$. 
By using Eqs.~(\ref{re1}) and
(\ref{re2}), we obtain $p$ and $\phi'^2p_X$ as the functions of $z$.
The energy density of the scalar field, $\rho=\dot{\phi}^2 p_X-p$,
is also known. {}From Eq.~(\ref{Idef}) we find 
\ba \label{Qevo}
Q=\frac{({\rm ln}\,I)'}{\phi'}\,. 
\ea
Hence the coupling $Q$ is determined once $I$ and $\phi'$ are known.
We need to specify the Lagrangian density $p(\phi, X)$
to find the evolution of $\phi'$ and $Q$.

The equation of state for dark energy, $w \equiv p/\rho$, is given by
\ba \label{w1}
w &=& \frac{p}{\dot{\phi}^2p_X-p} \\
\label{w2}
&=& \frac{(1+z)r'-3r}
{3r-3\Omega_{0m}(1+z)^3I}\,.
 \ea
A normal scalar field corresponds to $w>-1$, which translates into
the condition $p_X>0$ by Eq.\,(\ref{w1}). 
{}From Eq.~(\ref{re2})
we find that this condition corresponds to
$r'>3\Omega_{0m}(1+z)^2I$, which can be also checked by
Eq.~(\ref{w2}). Meanwhile a phantom field corresponds to $p_X<0$
and $r'<3\Omega_{0m}(1+z)^2I$. 
As we already mentioned, the evolution of $I(z)$ 
is determined if $r$ and $\delta_m$ are
known observationally. 
Then the equation of state of dark energy is obtained by
Eq.~(\ref{w2}) without specifying the Lagrangian 
density $p(\phi, X)$. 
In the next section we shall apply our formula to several 
different forms of scalar-field Lagrangian.

\section{Application to specific cases}
\label{app}

Most of scalar-field dark energy models
can be classified into two classes: 
(A) $p=f(X)-V(\phi)$ and (B) $p=f(X)V(\phi)$. 
There are special cases in which 
cosmological scaling solutions exist. This corresponds to 
the Lagrangian density 
(C) $p=Xg(Xe^{\lambda \phi})$, where $\lambda$
is a constant and $g$ is an arbitrary function. 
In what follows we shall consider these classes of 
models separately.

\subsection{Case of $p=f(X)-V(\phi)$}

This case includes quintessence [$f(X)=X$] and 
a phantom field [$f(X)=-X$].
{}From Eq.~(\ref{re2}) we find
\ba 
\label{or1}
\phi'^2 f_X=
\frac{r'-3\Omega_{0m}(1+z)^2I}{r(1+z)}\,.
\ea
If we specify the function $f(X)$, 
the evolution of $\phi'(z)$ and $\phi(z)$ is known 
from $r(z)$ and $I(z)$.
By Eq.~(\ref{Qevo}) we find the coupling $Q$
in terms of $z$ and $\phi$. 
Finally Eq.~(\ref{re1}) gives
\ba 
\label{or2}
V=f+[3r-(1+z)r']H_0^2 \,.
\ea
The r.h.s. is determined as a function of $z$.
Since $z$ is expressed by the field $\phi$, 
one can find the potential $V(\phi)$ in
terms of $\phi$. In the case of quintessence
without a coupling $Q$, this 
was carried out by a number of 
authors \cite{HT,Sta,NC,GOZ,SRSS}.
We have generalized this to a more general 
Lagrangian density 
$p=f(X)-V(\phi)$ in the presence of the coupling $Q$.

\subsection{Case of $p=f(X)V(\phi)$}

This case includes k-essence \cite{Kes} and 
tachyon \cite{tachyon} \footnote{We note that 
a coupled dark energy scenario between k-essence 
field and dark matter
was recently studied in Ref.~\cite{Wei}.}. 
The pressure density of the form
\ba
\label{Kes}
p(\tilde{X}, \varphi)=K(\varphi) \tilde{X}
+L(\varphi) \tilde{X}^2\,,~~~
\tilde{X}=-(\nabla \varphi)^2/2\,,
\ea
is transformed to the Lagrangian density 
$p=f(X)V(\phi)$ with $f(X)=-X+X^2$ and 
$V(\phi)=K^2/L$ by field redefinitions:
\ba 
\label{fre}
\phi=\int^{\varphi} \rd \varphi \sqrt{\frac{L}{|K|}}\,,~~~
X=\frac{L}{|K|} \tilde{X} \,. 
\ea
We note that dilatonic ghost condensate model \cite{PT} corresponds to 
a choice $K(\varphi)=-1$ and $L(\varphi)=ce^{\lambda \varphi}$. 
In this case the potential of the scalar field has a dependence $V(\phi)
\propto \phi^{-2}$ \cite{La,Cope} by the field 
redefinitions (\ref{fre}).

For the Lagrangian density $p=f(X)V(\phi)$ we obtain the following 
relation from Eqs.~(\ref{re1}) and (\ref{re2}): 
\ba & & \label{Ke1} \phi'^2
\frac{f_X}{f}=\frac{r'-3\Omega_{0m}(1+z)^2I}
{r(1+z)[(1+z)r'-3r]H_0^2}\,, \\
\label{Ke2}
& & V=\frac{[(1+z)r'-3r]H_0^2}{f}\,. \ea
Once we specify the form of $f(X)$, one can determine the functions
$\phi'(z)$ and $\phi(z)$ from Eq.~(\ref{Ke1}). Then
we obtain the potential $V(\phi)$ from Eq.~(\ref{Ke2}).

\subsection{Case of $p=Xg(Xe^{\lambda \phi})$}

When we construct realistic dark energy models, scaling solutions
may play an important role to solve coincident problems. 
In this case the energy density of the scalar field is proportional to 
that of 
a barotropic fluid ($\rho \propto \rho_m$).
It was shown in Refs.~\cite{PT,TS} that the
existence of scaling solutions restricts the form of the scalar-field 
Lagrangian to be
\ba 
\label{scaling}
p=Xg(Xe^{\lambda \phi})\,.
\ea
We note that this property holds both for coupled and 
uncoupled models of dark energy.

For the Lagrangian density (\ref{scaling}) 
Eqs.~(\ref{re1}) and (\ref{re2}) give
\ba \label{sca1} & &
Y\frac{g_Y}{g}=\frac{6r-(1+z)r'-3\Omega_{0m}(1+z)^3I}
{2[(1+z)r'-3r]}\,, \\
\label{sca2} & & 
\phi'^2=\frac{2[(1+z)r'-3r]}{r(1+z)^2g}\,, \ea
where $Y \equiv Xe^{\lambda \phi}$. If we specify the functional
form of $g(Y)$, one can determine the function $Y=Y(z)$ from
Eq.~(\ref{sca1}). Then we find $\phi'(z)$ and $\phi(z)$ by
Eq.~(\ref{sca2}). The parameter $\lambda$ is known 
by the relation
$Y=(1/2)\dot{\phi}^2 e^{\lambda \phi}$.

It was shown in Refs.~\cite{PT,TS} that the quantity $Y$ is
constant along scaling solutions, in which case the l.h.s. of
Eq.~(\ref{sca1}) is constant. 
In the absence of the coupling $Q$,
i.e., $I=1$, accelerated expansion is not realized 
for scaling solutions.
This means that the r.h.s. of Eq.~(\ref{sca1})
should not be constant for $Q=0$
when we use observational data. If the
coupling $Q$ is present, there is a possibility to find a
situation in which the r.h.s. of Eq.~(\ref{sca1}) does
not vary.
This corresponds to the case where accelerated expansion
is realized by scaling attractors. Thus one can directly check whether 
the present acceleration originates from 
scaling solutions with a non-zero coupling $Q$
once we obtain accurate observational data 
of $r(z)$ and $\delta_m(z)$.

\section{Examples of reconstruction}
\label{example}

In this section we show concrete examples of reconstruction
for several dark energy models. We shall use the following
parametrization for the Hubble parameter \cite{SSSA,ASSS}
\ba 
\label{para}
r(x)=\Omega_{0m}x^3+A_0+A_1x+A_2x^2\,,
\ea
where $x \equiv 1+z$ and $A_0 =1-A_1-A_2-\Omega_{0m}$. 
This corresponds to the following expansion 
for dark energy 
\ba 
\rho=\rho_{0c} \left(A_0+A_1x+A_2x^2\right)\,,
\ea
where $\rho_{0c}=3H_0^2$.

For a prior $\Omega_{0m}=0.3$, the
Gold dataset of SN observations gives
$A_1=-4.16 \pm 2.53$ and $A_2=1.67 \pm
1.03$ \cite{LNP}. We note that the weak energy condition 
for dark energy, $\rho \ge 0$ and $w=p/\rho \ge -1$, 
corresponds to \cite{ASSS}
\ba \label{weak}
A_0+A_1x+A_2x^2 \ge 0\,,
~~~~A_1+2A_2x \ge 0\,.
\ea
If we use the best-fit values $A_1=-4.16$ and
$A_2=1.67$, for example, we find that 
the second condition in Eq.~(\ref{weak})
is violated around present ($x \sim 1$). 
This means that the field
behaves as a phantom ($w<-1$). In the case of an ordinary 
scalar field such as quintessence, 
we need to put a prior $A_1+2A_2 x \ge 0$.

The reconstruction program of a quintessence-type 
scalar field has been already carried out 
in Ref.\,\cite{SRSS}, so we do not repeat it
here. We shall study the cases of tachyon and 
generalized ghost condensate when the coupling $Q$ is zero. 
We have not yet obtained good dataset about 
$\delta_m(z)$, so the coupling $Q$ is not well 
determined by current observations.

\subsection{Tachyon}

Tachyon corresponds to the case $f=-(1-2X)^{1/2}$ in
Eqs.~(\ref{Ke1}) and (\ref{Ke2}). Then we have
\ba & & \phi'^2 =
\frac{r'-3\Omega_{0m}x^2I}
{H_0^2rx[rx(r'-3\Omega_{0m}x^2I)+3r-xr']}\,, \\ 
& &
V= \frac{(3r-xr')H_0^2}
{\sqrt{1-r^2x^2H_0^2\phi'^2}}\,.
\ea
We note that the reconstruction equations were derived in 
Ref.~\cite{Hao} for $Q=0$.
The equation of state for tachyon is $w=\dot{\phi}^2-1$, 
which means that $w \ge -1$. 
Hence we should impose the prior given by
Eq.~(\ref{weak}).

\begin{figure}
\begin{center}
\includegraphics[height=3.2in,width=3.2in]{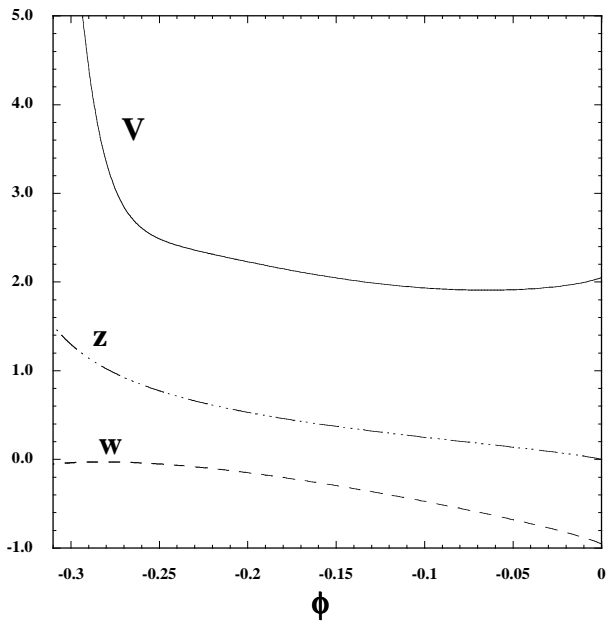}
\caption{\label{tachyon} 
Reconstruction of tachyon model for $Q=0$
for the parametrisation (\ref{para})
with $A_1=-3.80$ and $A_2=1.95$.  
These coefficients satisfy the 
weak energy condition (\ref{weak}), which means that the 
equation of state ranges in the region $w>-1$.
We show $V$, $w$ and $z$ in terms of the function of $\phi$.
We note that the potential is normalised by $H_0^2m_{\rm pl}^2$, 
where $m_{\rm pl}$ is the Planck mass.
}
\end{center}
\end{figure}

In Fig.~\ref{tachyon} we show one example of the reconstruction of
the tachyon model for $Q=0$ ($I=1$).
The field value is chosen to be $\phi=0$ at present ($z=0$).
The tachyon potential needs to be flat 
around $0<z<1$ to give rise to an accelerated expansion. 
The potential has a minimum 
with a non-zero energy density
for the parametrisation given 
in this figure ($A_1=-3.80$ and $A_2=1.95$).
This implies that the rolling massive scalar 
field model with potential 
$V(\phi)=V_0\exp(m^2\phi^2/2)$ \cite{GST}
can be a viable dark energy model.

\subsection{Generalised ghost condensate}

When the field satisfies the condition $p_X>0$, we need to 
put the prior (\ref{weak}) for consistency. Let us consider a
situation in which crossing of the cosmological-constant 
boundary is possible.
This can be realised for the following type of 
Lagrangian density:
\ba 
\label{gghost}
p=-X+h(\phi)X^2\,,
\ea
where $h(\phi)$ is a function in terms of $\phi$.
Dilatonic ghost condensate model \cite{PT} 
corresponds to a choice $h(\phi)=ce^{\lambda \phi}$. 
{}From Eqs.~(\ref{re1}) and (\ref{re2}) we obtain
\ba & & \phi'^2
=\frac{12r-3xr'-3\Omega_{0m}x^3I}{rx^2}\,, \\
& & h(\phi)=\frac{2(2xr'-6r+rx^2\phi'^2)}
{H_0^2r^2x^4 \phi'^4}\,.
 \ea

\begin{figure}
\begin{center}
\includegraphics[height=3.2in,width=3.2in]{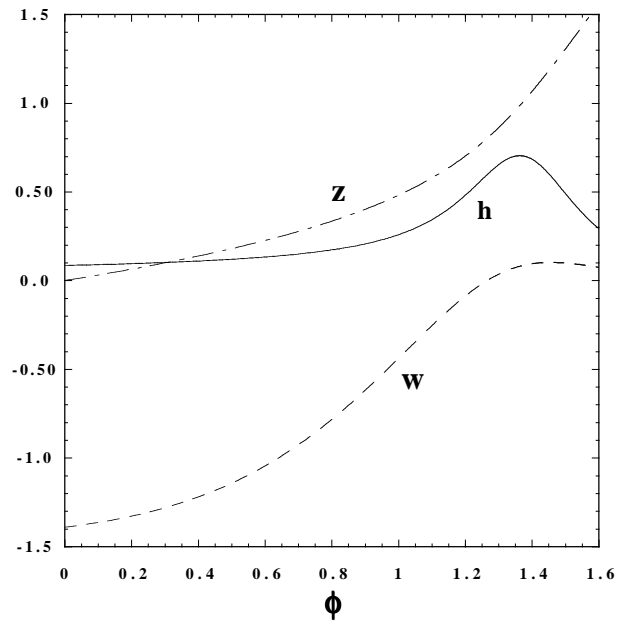}
\caption{\label{geghost} 
Reconstruction of generalized ghost condensate model 
for the parametrization (\ref{para})
with the best-fit parameters $A_1=-4.16$ and
$A_2=1.67$. 
We show $h$, $w$ and $z$ in terms of the function of $\phi$.
This model allows a possibility to cross the 
cosmological-constant boundary ($w=-1$).
}
\end{center}
\end{figure}

In Fig.\,\ref{geghost} we plot $h(\phi)$ in terms of the function 
$\phi$ when we use the best-fit values of $A_1$ and $A_2$. 
The crossing of the cosmological-constant boundary 
corresponds to $hX=1/2$, which 
occurs around the redshift $z=0.24$ for the best-fit
parametrisation. The system can enter the phantom region 
($hX<1/2$) without discontinuous behaviour of
$h$ and $X$. 

However we have to caution that the perturbation of the field $\phi$
is plagued by a quantum instability whenever it behaves 
as a phantom \cite{PT}. Even at the classical level
the perturbation is  unstable for $1/6<hX<1/2$, since
the speed of sound, $c_s^2=p_X/(p_X+2Xp_{XX})$,
becomes negative. One may avoid this instability
if the phantom behaviour is just transient.
In fact transient phantom behavior was found 
in the case of dilatonic ghost condensate model 
(see, e.g., Fig.\,4 in Ref.\,\cite{PT}). 
In this case the cosmological-constant boundary crossing
occurs again in future, after which the perturbations 
become stable.
 
We found that the function $h(\phi)$ can be approximated by 
an exponential function $e^{\lambda \phi}$ near to the present, 
although some difference appears for $z \gtrsim 0.2$.
However the current observational dataset is not still 
sufficient to rule out the dilatonic ghost condensate model.
We hope that future high-precision observations will 
determine the functional form of $h(\phi)$ more
accurately.

%
\subsection{Scaling solutions}

As we already mentioned, the existence of scaling solutions 
can be found by evaluating the r.h.s. of Eq.~(\ref{sca1}). 
When $Q=0$ we checked that the r.h.s. of Eq.~(\ref{sca1})
is not constant when we use the Gold dataset with the 
parametrisation (\ref{para}), which is 
consistent with the fact that scaling solutions do not lead to 
an accelerated expansion for $Q=0$.

In the presence of the coupling $Q$ if the solution around $0<z<1$
corresponds to a scaling solution, the r.h.s. of Eq.~(\ref{sca1})
is constant. This gives the constraint for the evolution of $I$:
\ba 
\label{Iz}
I(z)=\frac{r(r_0'-3\Omega_{0m})-xr'(1-\Omega_{0m})}
{\Omega_{0m} (r_0'-3)x^3}\,, 
\ea
where $I(0)=1$. 
The r.h.s. of this equation is independent of the scalar-field 
Lagrangian.

If both the evolution of $r(z)$ and $\delta_m(z)$ are known, 
one can determine $I(z)$ by Eq.~(\ref{Iprime}). 
Then by comparing this with Eq.~(\ref{Iz}), one can check 
the existence of scaling solutions.
It was shown in Ref.\,\cite{BNST} that in the case of 
a phantom field the final attractor does not correspond to 
scaling solutions but to scalar-field dominant solutions
with $\Omega_\phi=1$. 
Hence we have to caution that 
Eq.~(\ref{Iz}) can be used for the region 
characterised by Eq.~(\ref{weak}).

\section{Relationship with generalised Einstein theories}
\label{rela}

Having considered the coupling $Q$ between dark energy 
and dark matter, we would like to relate this scenario with
theories giving rise to such a coupling. 
Let us consider the following 4-dimensional 
Lagrangian density with a scalar field $\phi$
and a barotropic perfect fluid:
\ba
\label{stlag}
\tilde{{\cal L}}=\frac12  F(\vp) \tilde{R}-
\frac{1}{2}\zeta(\vp) (\tilde{\nabla} 
\vp)^2-U(\vp)-\tilde{{\cal L}}_m\,,
\ea
where $F(\vp)$, $\zeta(\vp)$ and $U(\vp)$ are 
the functions in terms of $\vp$. 
This includes a wide variety of gravity models--such 
as Brans-Dicke theories,
non-minimally coupled scalar fields and dilaton 
gravity\footnote{We note that the Lagrangian (\ref{stlag})
can be extended to the case in which higher-order curvature
corrections are taken into account. See Refs.~\cite{higher}
for details.}.

Making a conformal transformation $g_{\mu\nu}=
F(\vp)\tilde{g}_{\mu\nu}$, the above action 
reduces to that of the Einstein frame \cite{Maeda}:
\ba 
{\cal L}=\frac{1}{2}R-
\frac12(\nabla \phi)^2-V(\phi)
-{\cal L}_{m}(\phi)\,,
\ea
where 
\ba 
\label{phire}
\phi \equiv \int G(\vp)\rd \vp\,,~~~
G(\vp) \equiv \sqrt{\frac32
\left(\frac{F_\vp}{F}\right)^2+\frac{\zeta}{F}}\,.
\ea
We note that several quantities in Einstein frame are
related with those in string frame via
$a=\sqrt{F}\tilde{a}$, $\rd t=\sqrt{F}
\rd \tilde{t}$, $\rho_m=\tilde{\rho}_m/F^2$, 
$p_m=\tilde{p}_m/F^2$ and 
$V=U/F^2$. In Einstein frame the
background equations are given by Eqs.~(\ref{basiceq1}) and 
(\ref{basiceq2}) with $p=X-V(\phi)$ and
\ba \dot{\rho}_m+3H(\rho_m+p_m)=
-\frac{F_\vp}{2FG}(\rho_m-3p_m)\dot{\phi}\,.
\ea
In the case of non-relativistic dark matter ($p_m=0$) this
corresponds to Eq.~(\ref{basiceq3}) with a coupling
\ba 
\label{coup}
Q(\vp)=-\frac{F_\vp}{2F} \left[\frac32
\left(\frac{F_\vp}{F}\right)^2+
\frac{\zeta}{F}\right]^{-1/2}\,.
\ea

For example a nonminimally coupled scalar field with a 
coupling $\xi$ corresponds to
$F(\vp)=1-\xi \vp^2$ and $\zeta(\vp)=1$. 
In this case we find
\ba 
\label{Qxi}
Q(\vp)=\frac{\xi \vp}{[1-\xi \vp^2(1-6\xi)]^{1/2}}\,.
\ea
Then $\xi$ is expressed in terms of $Q$:
\ba 
\label{xievo}
\xi=\frac{Q^2}{2(1-6Q^2)} \left[-1 \pm
\sqrt{1+\frac{4(1-6Q^2)}
{Q^2\vp^2}}\right]\,. 
\ea
Once we know $r(z)$ and $\delta_m(z)$, 
the coupling $\xi$ is evaluated by using 
Eqs.~(\ref{Qevo}) and (\ref{xievo}) 
together with the relation (\ref{phire}) between 
$\phi$ and $\varphi$.
Hence it is possible to determine the strength of the 
nonminimal coupling from observations.
We note that $Q \simeq \xi \vp$ for $|\xi| \ll 1$
and $Q \simeq \pm 1\sqrt{6}$ in the large-coupling 
limit ($|\xi| \gg 1$). 
The large-coupling case is excluded from solar system 
experiments provided that the scalar field is universally 
coupled to all matter \cite{Lucanon,Luca}.
It is certainly of interest to place constraints on the strength of $\xi$
by using our reconstruction formula together with other experiments
about the time-variation of a gravitational ``constant'' $G$.

String theory also gives rise to the coupling $Q$ after a conformal 
transformation from string frame to Einstein frame. 
The tree-level dilaton gravity \cite{Gas} corresponds to 
$F(\vp) \sim e^{-\vp}$ and $\zeta(\vp) \sim -e^{-\vp}$, which gives
a constant value of $Q$ by Eq.~(\ref{coup}). 
It is typically assumed that nonperturbative
effects would stabilize the dilaton with a potential so that 
it does not contradict with solar system experiments.
An alternative possibility is the runaway dilaton 
scenario \cite{runaway} in which the dilaton is effectively 
decoupled from gravity in the limit 
$\varphi \to \infty$ with the dependence $F(\vp) \sim B_1+C_1e^{-\vp}$
and $\zeta(\vp) \sim B_2+C_2e^{-\vp}$. 
One can also check the viability of this scenario
by comparing Eq.~(\ref{coup}) with the coupling $Q$ obtained 
from observations.

There is another interesting cosmological scenario in which 
neutrinos are coupled to dark energy \cite{neutrino}.
In this model the neutrino mass, $m_\nu$, is a function of
a scalar field, $\phi$.
In a situation where neutrinos are collisionless, the neutrino 
energy density, $\rho_\nu$, satisfies the 
equation of motion \cite{neutrino}
\ba 
\label{neu}
\dot{\rho}_\nu+3H(\rho_\nu+p_\nu)=
\frac{\partial {\rm ln}\,m_\nu}{\partial \phi}
(\rho_\nu-3p_\nu)\dot{\phi}\,,
\ea
where $p_\nu$ is the pressure density of neutrinos.
When the neutrinos become non-relativistic, Eq.~(\ref{neu})
shows that the coupling between neutrinos and dark energy is 
given by $Q(\phi)=\partial {\rm ln}\, m_\nu/\partial \phi$.
Hence if we know the coupling $Q(\phi)$ observationally,  
the evolution of the neutrino mass is found as a function of 
$\phi$ (and $z$).

\section{Conclusions}

In this paper we have provided a method to reconstruct 
scalar-field Lagrangian in an accelerating universe from observations.
Our starting point is the general Lagrangian (\ref{action})
which is the function of a scalar-field $\phi$ 
and a kinematic term $X$.
We have also taken into account the coupling $Q(\phi)$ 
between dark energy and a non-relativistic perfect fluid
in order to include the coupled quintessence scenario.

In the absence of the coupling $Q$, one can reconstruct the 
structure of theory by parametrising the Hubble rate $H$
in terms of the redshift $z$ from the luminosity distance of 
supernovae observational data. We need to know additional 
information in order to determine the coupling $Q$ from 
observations. We have made use of the equation for matter
density perturbations $\delta_m$ on sub-Hubble scales
for this purpose. Our reconstruction formula is given 
by Eqs.~(\ref{re1}), (\ref{re2}) and (\ref{Iprime}), 
from which the coupling $Q$ is also determined
with the use of Eq.~(\ref{Qevo}).
We note that the equation of of state (\ref{w2}) for dark energy 
is known from observables only without 
specifying any Lagrangian.

In Sec.~\ref{app} we have applied our reconstruction formula for several 
forms of Lagrangian density: (i) $p=f(X)-V(\phi)$, (ii) $p=f(X)V(\phi)$
and (iii) $p=Xg(Xe^{\lambda \phi})$ where $g$ is an arbitrary function.
In the cases (i) and (ii) reconstruction equations can be 
decomposed into two contributions coming from a kinematic 
term and a potential term.
Hence the potentials of such theories can be obtained together 
with the coupling $Q$ once the evolution of 
$H(z)$ and $\delta_m (z)$ is known. 
The case (iii) corresponds to the Lagrangian density 
for the existence of scaling solutions.
One can check the existence of scaling solutions 
if we evaluate the r.h.s. of Eq.~(\ref{sca1}) using 
observational data.

In Sec.~\ref{example} we have presented concrete examples of
our reconstruction with the parametrization given by 
Eq.~(\ref{para}). We studied two dark energy scenarios in the 
absence of the coupling $Q$-- (a) tachyon  and (b) generalised
ghost condensate. In the case (a) the equation of state for the tachyon 
field is constrained to be $w>-1$, which means that the prior
(\ref{weak}) needs to be imposed. 
In Fig.~\ref{tachyon} we plotted one 
example for the reconstruction of the tachyon potential.
The model (b) allows a possibility to cross the 
cosmological-constant boundary: $w=-1$. 
We carried out the reconstruction of 
this model by using the best-fit values coming from  
the Gold dataset. The result is illustrated in Fig.~\ref{geghost},
which shows that the field behaves as a phantom 
for the redshift $0<z<0.24$.

In Sec.~\ref{rela} we presented a Lagrangian in generalised 
Einstein theories which gives rise to the coupling $Q$
by a conformal transformation to Einstein frame. 
For example, nonminimal 
coupling $\xi$ is directly related with $Q$ 
as given in Eq.~(\ref{xievo}). This allows a possibility to determine 
the strength of the nonminimal coupling from observational 
data of $H(z)$ and $\delta_m(z)$.

For the moment we have not yet obtained the accurate evolution 
of $\delta_m(z)$ from observations of clustering.
We only know the total amount of growth between the decoupling 
epoch and present. This is associated with the fact that all probes of 
clustering are plagued by a bias problem.
However upcoming galaxy surveys such as  KAOS, LSST
and PANSTARS will pin down the matter power spectrum to exquisite
accuracy, allowing the ultimate measurement of the power spectrum.
By that time we should have an excellent understanding of bias
and will be able to obtain the time-evolution of $\delta_m$.
We hope that this will provide us an exciting possibility 
to reveal the origin of dark energy.

\section*{ACKNOWLEDGMENTS}
The author thanks Luca Amendola, M.~Sami and Alexander
Vikman for useful discussions. He is also grateful to 
Martin Bojowald, Mariusz Dabrowski and Burin Gumjudpai for supporting
visits to Max Planck institute, Pomeranian Workshop and 
Naresuan University. This work is supported by JSPS
(Grant No.\,30318802).



\begin{thebibliography}{40}

\bibitem{Perl}
A.~G.~Riess {\it et al.}  [Supernova Search Team Collaboration],
Astron.\ J.\  {\bf 116}, 1009 (1998);
S.~Perlmutter {\it et al.},
Astrophys.\ J.\  {\bf 517}, 565 (1999).

\bibitem{SN}
A.~G.~Riess {\it et al.},
Astrophys.\ J.\  {\bf 607}, 665 (2004).

\bibitem{CMB}
D.~N.~Spergel {\it et al.},
Astrophys.\ J.\ Suppl.\  {\bf 148}, 175 (2003).

\bibitem{SDSS}
M.~Tegmark {\it et al.},
Phys.\ Rev.\ D {\bf 69}, 103501 (2004); 
Astrophys.\ J.\  {\bf606}, 702 (2004).

\bibitem{obser}
S.~Hannestad and E.~Mortsell,
Phys.\ Rev.\ D {\bf 66}, 063508 (2002);
A.~Melchiorri, L.~Mersini, C.~J.~Odman and M.~Trodden,
Phys.\ Rev.\ D {\bf 68}, 043509 (2003);
J.~Weller and A.~M.~Lewis,
Mon.\ Not.\ Roy.\ Astron.\ Soc.\  {\bf 346}, 987 (2003);
Y.~Wang and M.~Tegmark,
Phys.\ Rev.\ Lett.\  {\bf 92}, 241302 (2004);
B.~Feng, X.~L.~Wang and X.~M.~Zhang,
Phys.\ Lett.\ B {\bf 607}, 35 (2005);
T.~R.~Choudhury and  T.~Padmanabhan, 
Astron. Astrophys. {\bf 429}, 807 (2005);
H.~K.~Jassal, J.~S.~Bagla and T.~Padmanabhan,
Mon.\ Not.\ Roy.\ Astron.\ Soc.\  {\bf 356}, L11 (2005);
P.~S.~Corasaniti, M.~Kunz, D.~Parkinson, E.~J.~Copeland
and B.~A.~Bassett,
Phys.\ Rev.\ D {\bf 70}, 083006 (2004);
U.~Seljak {\it et al.},
Phys.\ Rev.\ D {\bf 71}, 103515 (2005);
M.~Sahlen, A.~R.~Liddle and D.~Parkinson,
arXiv:astro-ph/0506696.

\bibitem{review}
V.~Sahni and A.~A.~Starobinsky,
Int.\ J.\ Mod.\ Phys.\ D {\bf 9}, 373 (2000); 
V.~Sahni,
arXiv:astro-ph/0403324;
T.~Padmanabhan,
Phys.\ Rept.\  {\bf 380}, 235 (2003).

\bibitem{quin}
C.~Wetterich, Nucl. Phys. {\bf B302}, 668 (1988); 
B.~Ratra and P.~J.~E.~Peebles, Phys. Rev. D{\bf 37}, 
3406 (1988); 
E.~J.~Copeland, A.~R.~Liddle, and D.~Wands, 
Ann. N. Y. Acad. Sci. {\bf 688}, 647 (1993); 
P.~G.~Ferreira and M.~Joyce, Phys. Rev. Lett. {\bf 79}, 
4740 (1997); Phys. Rev D {\bf 58}, 023503 (1998);
I.~Zlatev, L.~M.~Wang and P.~J.~Steinhardt,
Phys.\ Rev.\ Lett.\  {\bf 82}, 896 (1999); P.~J.~Steinhardt,
L.~M.~Wang and I.~Zlatev,
Phys.\ Rev.\ D {\bf 59}, 123504 (1999).

\bibitem{Kes}
C.~Armendariz-Picon, V.~Mukhanov and P.~J.~Steinhardt,
Phys.\ Rev.\ Lett.\  {\bf 85}, 4438 (2000); Phys.\ Rev.\ D {\bf
63}, 103510 (2001); T.~Chiba, T.~Okabe and M.~Yamaguchi,
Phys.\ Rev.\ D {\bf 62}, 023511 (2000).

\bibitem{tachyon}
G.~W.~Gibbons,
Phys.\ Lett.\ B {\bf 537}, 1 (2002); T.~Padmanabhan,
Phys.\ Rev.\ D {\bf 66}, 021301 (2002); J.~S.~Bagla, H.~K.~Jassal
and T.~Padmanabhan,
Phys.\ Rev.\ D {\bf 67}, 063504 (2003).

\bibitem{phantom}
R.~R.~Caldwell, Phys.\ Lett.\ B {\bf 545}, 23 (2002);
R.~R.~Caldwell, M.~Kamionkowski and N.~N.~Weinberg,
Phys.\ Rev.\ Lett.\  {\bf 91}, 071301 (2003);
P.~Singh, M.~Sami and N.~Dadhich,
Phys.\ Rev.\ D {\bf 68}, 023522 (2003).

\bibitem{Arkani}
N.~Arkani-Hamed, H.~C.~Cheng, M.~A.~Luty and S.~Mukohyama,
JHEP {\bf 0405}, 074 (2004).

\bibitem{PT}
F.~Piazza and S.~Tsujikawa,
JCAP {\bf 0407}, 004 (2004).

\bibitem{KN}
A.~Krause and S.~P.~Ng,
arXiv:hep-th/0409241.

\bibitem{Sta}
A.~A.~Starobinsky, JETP Lett.\  {\bf 68}, 757 (1998) 
[Pisma Zh.\ Eksp.\ Teor.\ Fiz.\  {\bf 68}, 721 (1998)].

\bibitem{HT}
D.~Huterer and M.~S.~Turner,
Phys.\ Rev.\ D {\bf 60}, 081301 (1999).

\bibitem{NC}
T.~Nakamura and T.~Chiba,
Mon.\ Not.\ Roy.\ Astron.\ Soc.\  {\bf 306}, 
696 (1999).

\bibitem{GOZ}
Z.~K.~Guo, N.~Ohta and Y.~Z.~Zhang,
Phys.\ Rev.\ D {\bf 72}, 023504 (2005).

\bibitem{SRSS}
T.~D.~Saini, S.~Raychaudhury, V.~Sahni
and A.~A.~Starobinsky,
Phys.\ Rev.\ Lett.\  {\bf 85}, 1162 (2000).

\bibitem{BEPS}
B.~Boisseau, G.~Esposito-Farese, D.~Polarski and
A.~A.~Starobinsky,
Phys.\ Rev.\ Lett.\  {\bf 85}, 2236 (2000).

\bibitem{EP}
G.~Esposito-Farese and D.~Polarski,
Phys.\ Rev.\ D {\bf 63}, 063504 (2001).

\bibitem{Pe}
L.~Perivolaropoulos,
arXiv:astro-ph/0504582.

\bibitem{Capo}
S.~Capozziello, V.~F.~Cardone and A.~Troisi,
Phys.\ Rev.\ D {\bf 71}, 043503 (2005).

\bibitem{CCENO}
S.~Capozziello, V.~F.~Cardone, E.~Elizalde, 
S.~Nojiri and S.~D.~Odintsov,
arXiv:astro-ph/0508350. \\
See also \\
I.~Brevik and O.~Gorbunova,
arXiv:gr-qc/0504001;
S.~Nojiri and S.~D.~Odintsov,
Phys.\ Rev.\ D {\bf 72}, 023003 (2005).

\bibitem{Luca}
L.~Amendola,
Phys.\ Rev.\ D {\bf 62}, 043511 (2000).

\bibitem{Lucanon}
L.~Amendola,
Phys.\ Rev.\ D {\bf 60}, 043501 (1999).

\bibitem{Uzan}
J.~P.~Uzan,
Phys.\ Rev.\ D {\bf 59}, 123510 (1999).

\bibitem{Tchiba}
T.~Chiba,
Phys.\ Rev.\ D {\bf 60}, 083508 (1999).

\bibitem{BP}
N.~Bartolo and M.~Pietroni,
Phys.\ Rev.\ D {\bf 61} 023518 (2000).

\bibitem{Mata}
F.~Perrotta, C.~Baccigalupi and S.~Matarrese,
Phys.\ Rev.\ D {\bf 61}, 023507 (2000).

\bibitem{Amenphan}
L.~Amendola,
Phys.\ Rev.\ Lett.\  {\bf 93} 181102 (2004).

\bibitem{Amenper}
L.~Amendola, Phys.\ Rev.\ D {\bf 70}, 103524 (2004).

\bibitem{ATS}
L.~Amendola, S.~Tsujikawa and M.~Sami,
arXiv:astro-ph/0506222.

\bibitem{gala1}
W.~J.~Percival {\it et al.},  
MNRAS {\bf 327}, 1297 (2001).

\bibitem{gala2}
M.~Tegmark {\it et al.},  
APJ {\bf 606}, 702 (2004).

\bibitem{gala3}
A.~C.~Pope {\it et al.},  
APJ {\bf 607}, 655 (2004).

\bibitem{Sel1}
U.~Seljak {\it et al.}, astro-ph/0406594.

\bibitem{Sel2}
U.~Seljak {\it et al.}, astro-ph/0407372.

\bibitem{Vikman}
A.~Vikman,
Phys.\ Rev.\ D {\bf 71}, 023515 (2005);
A.~Anisimov, E.~Babichev and A.~Vikman,
JCAP {\bf 0506}, 006 (2005).

\bibitem{NOT}
S.~Nojiri, S.~D.~Odintsov and S.~Tsujikawa,
Phys.\ Rev.\ D {\bf 71}, 063004 (2005).

\bibitem{multi}
Z.-K. Guo, Y.-S. Piao, X. Zhang, and Y.-Z. Zhang, 
Phys.\ Lett.\ B {\bf 608}, 177 (2005);
W.~Hu, Phys.\ Rev.\ D {\bf 71}, 047301 (2005);
H.~Wei, R.~G.~Cai and D.~F.~Zeng,
Class.\ Quant.\ Grav.\  {\bf 22}, 3189 (2005);
H.~Stefancic,
Phys.\ Rev.\ D {\bf 71}, 124036 (2005);
R.~R.~Caldwell and M.~Doran,
arXiv:astro-ph/0501104;
X.~F.~Zhang, H.~Li, Y.~S.~Piao and X.~M.~Zhang,
arXiv:astro-ph/0501652;
P.~Singh,
arXiv:gr-qc/0502086;
M.~z.~Li, B.~Feng and X.~m.~Zhang,
arXiv:hep-ph/0503268.

\bibitem{Wei}
H.~Wei and R.~G.~Cai,
Phys.\ Rev.\ D {\bf 71}, 043504 (2005).

\bibitem{La}
J.~M.~Aguirregabiria and R.~Lazkoz,
Phys.\ Rev.\ D {\bf 69}, 123502 (2004).

\bibitem{Cope}
E.~J.~Copeland, M.~R.~Garousi, M.~Sami and S.~Tsujikawa,
Phys.\ Rev.\ D {\bf 71}, 043003 (2005).

\bibitem{TS}
S.~Tsujikawa and M.~Sami, 
Phys.\ Lett.\ B \textbf{603}, 113 (2004).

\bibitem{SSSA}
V.~Sahni, T.~D.~Saini, A.~A.~Starobinsky and U.~Alam,
JETP Lett.\  {\bf 77}, 201 (2003) [Pisma Zh.\ Eksp.\ Teor.\ Fiz.\
{\bf 77}, 249 (2003)].

\bibitem{ASSS}
U.~Alam, V.~Sahni, T.~D.~Saini and A.~A.~Starobinsky,
Mon.\ Not.\ Roy.\ Astron.\ Soc.\  {\bf 354}, 275 (2004).

\bibitem{LNP}
R.~Lazkoz, S.~Nesseris and L.~Perivolaropoulos,
arXiv:astro-ph/0503230.

\bibitem{Hao}
J.~g.~Hao and X.~z.~Li,
Phys.\ Rev.\ D {\bf 66}, 087301 (2002).

\bibitem{GST}
M.~R.~Garousi, M.~Sami and S.~Tsujikawa,
Phys.\ Rev.\ D {\bf 70}, 043536 (2004);
Phys.\ Lett.\ B {\bf 606}, 1 (2005).

\bibitem{BNST}
B.~Gumjudpai, T.~Naskar, M.~Sami and S.~Tsujikawa,
JCAP {\bf 0506}, 007 (2005).
 
\bibitem{higher}
E.~Elizalde, S.~Nojiri and S.~D.~Odintsov,
Phys.\ Rev.\ D {\bf 70}, 043539 (2004);
S.~Nojiri, S.~D.~Odintsov and M.~Sasaki,
Phys.\ Rev.\ D {\bf 71}, 123509 (2005);
M.~Sami, A.~Toporensky, P.~V.~Tretjakov 
and S.~Tsujikawa,
Phys.\ Lett.\ B {\bf 619}, 193 (2005);
G.~Calcagni, S.~Tsujikawa and M.~Sami,
arXiv:hep-th/0505193.
 
\bibitem{Maeda}
K.~i.~Maeda,
Phys.\ Rev.\ D {\bf 39}, 3159 (1989).

\bibitem{Gas}
G.~Veneziano,
Phys.\ Lett.\ B {\bf 265}, 287 (1991);
M.~Gasperini and G.~Veneziano,
Astropart.\ Phys.\  {\bf 1} 317 (1993).

\bibitem{runaway}
M.~Gasperini, F.~Piazza and G.~Veneziano,
Phys.\ Rev.\ D {\bf 65} 023508 (2002);
T.~Damour, F.~Piazza and G.~Veneziano,
Phys.\ Rev.\ Lett.\  {\bf 89}, 081601 (2002).

\bibitem{neutrino}
R.~Fardon, A.~E.~Nelson and N.~Weiner, 
JCAP {\bf 0410}, 005 (2004);
X.-J.~Bi, P.~Gu, X.~Wang and X.~Zhang, 
Phys.\ Rev.\ D {\bf 69} 113007 (2004);
P.~Hung and H.~Das, astro-ph/0311131;
D.~B. Kaplhan, A.~E.~Nelson and N.~Weiner, 
Phys.\ Rev.\ Lett.\  {\bf 93}, 091801 (2004);
R.~D.~Peccei, Phys.\ Rev.\ D {\bf 71} 023527 (2005);
E.~I.~Guendelman and A.~B.~Kaganovich,
hep-th/0411188;
X.~Bi, B.~Feng, H.~Li and X. Zhang, hep-th/0412002;
V.~Barger, P.~Huber and D.~Marfatia, hep-ph/0502196;
M.~Cirelli, M.~C.~Gonzalez-Garcia and C.~Pena-Garay,
hep-ph/0503028;
A.~W.~Brookfield, C.~van de Bruck, D.~F.~Mota
and D.~Tocchini-Valentini, astro-ph/0503349.

\end{thebibliography}
\end{document}